\documentclass[aps,prl,reprint,showpacs,superscriptaddress,nobibnotes]{revtex4-2}
\pdfoutput=1
\usepackage{graphicx}
\usepackage{color}
\usepackage{orcidlink}
\usepackage{amsmath}
\usepackage{amssymb}
\usepackage{physics}
\usepackage[caption=false]{subfig}
\usepackage{slashed} 
\usepackage{ulem}
\usepackage{hyperref}
\newcommand{\AppOne}{Appendix~A}
\newcommand{\AppTwo}{Appendix~B}

\begin{document}

\title{
The Collins-Soper kernel from a vacuum soft function
}
\author{Anthony~Francis\orcidlink{0000-0002-3303-9900}}
\email{afrancis@nycu.edu.tw}
\affiliation{Institute of Physics, National Yang Ming Chiao Tung
  University, Hsinchu 30010, Taiwan}

\author{C.-J.~David~Lin\orcidlink{0000-0003-3743-0840}}
\email{dlin@nycu.edu.tw}
\affiliation{Institute of Physics, National Yang Ming Chiao Tung
  University, Hsinchu 30010, Taiwan}
\affiliation{Centre for High Energy Physics, Chung-Yuan Christian
  University, Taoyuan 32023, Taiwan}
\affiliation{Physics Division, National Centre for Theoretical 
  Sciences, Taipei 106319, Taiwan}

\author{Wayne~Morris\orcidlink{0000-0003-2430-0772}}
\email{waynemorris@nycu.edu.tw}
\affiliation{Institute of Physics, National Yang Ming Chiao Tung
  University, Hsinchu 30010, Taiwan}

\author{Yong~Zhao\orcidlink{0000-0002-2688-6415}}
\email{yong.zhao@anl.gov}
\affiliation{Physics Division, Argonne National Laboratory, Lemont, IL 60439, USA}

\begin{abstract} 
The Collins-Soper kernel is calculated from a vacuum soft function
using space-like Wilson lines with complex-directional vectors on the Euclidean lattice.
Our pure gauge calculations with this method
achieve high statistical precision in computing the soft function, whose rapidity dependence is well described by Collins-Soper evolution across a wide range of rapidity differences.
The extracted kernel contains errors comparable to those achieved in state-of-the-art lattice calculations based on hadronic observables, but exhibits saturated behavior at large transverse Wilson-line separations.
\end{abstract}



\maketitle



\paragraph{Introduction.}

Parton distribution functions (PDFs) are central objects in the investigation of 
hadron structure, providing information on 
the distribution of strongly interacting matter therein.
The transverse degrees of freedom can be be described by the
transverse momentum dependent PDFs (TMDPDFs).
In theoretical calculations, the TMDPDF 
\cite{Collins:2011zzd,Collins:1981uk,Collins:1981va,Collins:1984kg,Collins:1985ue,Collins:1988ig,Catani:2000vq,Catani:2010pd,Collins:1989gx,deFlorian:2001zd}
is expressed in terms of the so-called soft and beam functions, computed from matrix elements involving Wilson lines along the light-cone.
Respectively, these objects contain information about
soft and collinear modes in scattering processes, 
such as Drell-Yan or semi-inclusive deeply inelastic scattering (SIDIS).
Both objects 
evolve with a rapidity scale, governed by the so-called Collins-Soper (CS) kernel, 
which plays a crucial role in
extracting TMDPDFs from experiments \cite{Angeles-Martinez:2015sea,Amoroso:2022eow}.

At small parton transverse momentum the CS kernel cannot be obtained in perturbation theory, but is instead extracted through fitting experimental data in the process of obtaining TMDPDFs \cite{Hautmann:2020cyp,Bury:2022czx,Bacchetta:2022awv,Isaacson:2023iui,Moos:2023yfa,Aslan:2024nqg,Bacchetta:2024yzl,bacchetta2025,Kang:2024dja,Martinez:2024mou,Moos:2025sal,Camarda:2025lbt,Barry:2025glq,Avkhadiev:2025wps,Aglietti:2026bhu,Kang:2026mod}.
These phenomenological fits require parametrization of the CS kernel, whose model uncertainty remains challenging to quantify,
motivating
first-principles, non-perturbative QCD calculations. 
So far, Large Momentum Effective Theory (LaMET)~\cite{Ji:2013dva,Ji:2014gla,Ji:2020ect} has provided a framework to calculate the TMDPDFs~\cite{Ji:2014hxa,Ji:2018hvs,Ebert:2018gzl,Ebert:2019okf,Ji:2019sxk,Ji:2019ewn,Ebert:2020gxr,Ji:2020jeb,Ji:2021znw,Ebert:2022fmh,Schindler:2022eva,Deng:2022gzi,Zhu:2022bja,delRio:2023pse,Ji:2023pba,Zhao:2023ptv,Xie:2025rrw} and the CS kernel~\cite{Ji:2014hxa,Ebert:2018gzl,Ji:2019sxk} from lattice QCD. In LaMET, the CS kernel can be extracted from the momentum evolution of a quasi-TMDPDF~\cite{Ebert:2018gzl} or quasi-TMD wave function~\cite{Ji:2019sxk,Ji:2021znw}, which are hadronic matrix elements of equal-time correlators. Progress has been made towards systematic lattice calculations of the kernel~\cite{Shanahan:2020zxr,LatticeParton:2020uhz,Schlemmer:2021aij,Shanahan:2021tst,Li:2021wvl,LatticePartonLPC:2022eev,LatticePartonLPC:2023pdv,Shu:2023cot,Avkhadiev:2023poz,Liu:2024sqj,Avkhadiev:2024mgd,Bollweg:2024zet,Bollweg:2025iol,Alexandrou:2025xci,Tan:2025ofx,Tan:2026ier}, notably with continuum extrapolations at physical quark masses~\cite{Avkhadiev:2023poz,Tan:2025ofx}. LaMET requires simulating hadrons at large momenta to suppress power corrections, which is the most computationally demanding part of the calculation and a major source of systematic uncertainty.

As a universal quantity in QCD, the CS kernel can also be extracted from the rapidity evolution of the soft function~\cite{Collins:2011zzd}.
On the lattice, this approach could remove the computational overhead for simulating highly boosted hadrons, thereby significantly improving statistical precision.
It was suggested in Ref.~\cite{Ji:2019sxk} that a soft function with time-like Wilson lines could be computed on the lattice by simulating a moving, infinite-mass, non-relativistic heavy quark~\cite{Horgan:2009ti,Aglietti:1993hf,Hashimoto:1995in}. We investigate this idea and find that only space-like Wilson lines reproduce the correct rapidity dependence of the soft function in Euclidean space. Furthermore, we develop a detailed strategy of its implementation on the lattice using the auxiliary-field representation of Wilson lines \cite{Arefeva:1980zd, Gervais:1979fv} with complex-directional vectors.

We present the first lattice calculation of the CS kernel through a vacuum soft function using this strategy. By adjusting the complex-directional vectors we are able to construct Wilson lines of arbitrary rapidity, which is not limited by the lattice cutoff like the hadron momentum. We achieve high statistical precision in computing the soft function on three pure SU(3) gauge ensembles with fine lattice spacings. Within a wide range of rapidities, the rapidity-dependence of the soft function is well described by CS evolution. 
We fix the renormalization of Wilson-line rapidity by matching lattice results to perturbation theory at short transverse separations, allowing us to determine the CS kernel in the non-perturbative region. 
Although uncertainties in the final results are dominated by perturbative matching systematics, they are comparable to state-of-the-art LaMET calculations~\cite{Avkhadiev:2024mgd,Tan:2025ofx}, demonstrating the predictive power of our approach. On pure gauge ensembles, we observe that the CS kernel approaches a constant at large transverse separations, which can have important implications for TMD phenomenology.

\paragraph{Theoretical framework.} 
Working in terms of the Fourier transformed
TMDPDF, 
$f_R^{\rm TMD}\left(x,b_\perp,\mu,\zeta\right)$, the CS kernel is defined as~\cite{Collins:2011zzd}:
\begin{align}
\gamma_R\left(b_\perp, \mu\right) 
    &= \dv{\sqrt{\zeta}} \ln f^{\rm TMD}_R \left(x, b_\perp, \mu, \zeta \right) \, ,
\end{align}
where $x$ is the parton momentum fraction, $b_\perp$ is the Fourier conjugate to the parton transverse momentum, $k_\perp$, $\mu$ is the renormalization scale, and
$\zeta = 2\left(x p^+ e^{-y_n}\right)^2$ is the CS scale, 
where 
$y_n$ is a rapidity parameter and $p^+ = p^0 + p^3$ is the hadron momentum. 
The subscript, $R\in\left\{Q,G\right\}$, represents a quark or gluon, respectively.
The CS scale
arises from regulating rapidity divergences in 
$f_R^{\rm TMD}$, the beam function, and the soft function, $S_R$.
The rapidity scale evolution of $S_R$ is: 
\begin{align}
\gamma_R \left(b_\perp, \mu\right) 
&= \dv{\ln S_R\left(b_\perp,\epsilon,y\right)}{y}
- {\rm UV~counter\text{-}terms} \, ,
\end{align}
allowing one to obtain the CS kernel from 
a computation of the soft function.

\begin{figure}[t]
    \includegraphics[scale=0.26]{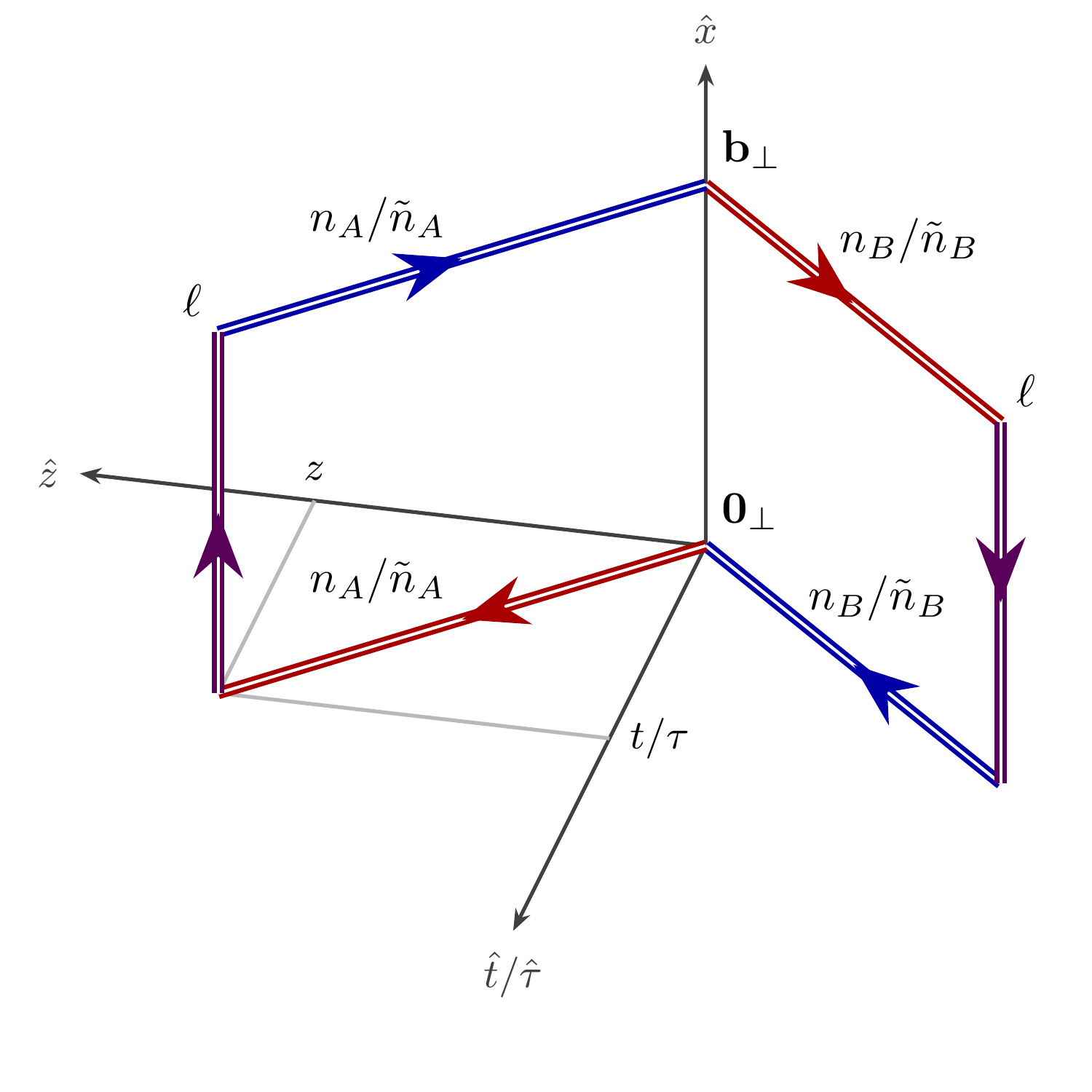}
    \caption{Schematic of the Wilson loops in Eqs. (\ref{eq:soft_def}), and (\ref{eq:bfly_op}).}
    \label{fig:bflyL}
\end{figure}

The soft function is defined as the vacuum expectation value
of two staple shaped products of Wilson lines that point along separate
light-cone directions, illustrated by 
Fig. \ref{fig:bflyL} in the limit $\ell\to\infty$, where $\ell$ is the staple length. 
We write the matrix element for the bare soft function as
\begin{align}
    \begin{split}
        &S_Q^{(0)}\left(b_\perp; \eta; \epsilon \right) \\
    & \quad = \frac 1{N_c}
    \bra{0}
        \Tr \bigl[
            W_n\left(b_\perp;0,-\infty\right)
            W_{\bar n}\left(b_\perp;-\infty,0\right) \bigr. \\
            & \qquad\qquad \times
            W_{\perp}\left(-\infty \,\bar n; 0, b_\perp\right)
            W_{\bar n}\left(0;0,-\infty\right) \\
            & \qquad\qquad \times \bigl. 
            W_{n}\left(0;-\infty,0\right)
            W_{\perp}\left(-\infty\, n; b_\perp, 0\right)
        \bigr]_\eta
    \ket{0} \, ,
    \end{split}
    \label{eq:soft_def}
\end{align}
where $\eta$ is a rapidity regulator, $\epsilon$ is an ultraviolet
(UV) regulator,  and the
Wilson line is written as:
\begin{align}
    W_{v}\!\left(x^\mu;a,b\right)
    &\!=\!
    P\exp \Big[
        \!-ig_0 
        \int_a^b\! \dd s v\cdot A \left(x^\mu + s v^\mu\right)
        \Big] \, ,
\end{align}
where $v=n/\bar n$ corresponds to light-cone directions, and $v=\hat b_\perp$ corresponds to a unit vector in the transverse direction.
The light-cone directions are defined by
$n,~\bar n$.
In Minkowski space, the soft function suffers from rapidity
divergences associated with 
Wilson lines pointing along the light-cone. One method for regulating such divergences is
off-light-cone regulators. In Collins' regularization scheme \cite{Collins:2011zzd}, one deforms the 
directional vectors into the space-like region,
\begin{align}
    n_{A/B} &= \Big( ~1-e^{\mp 2y_{A/B}},~0, ~0, ~\pm (1 + e^{\mp 2y_{A/B}})  ~\Big) \, ,
\end{align}
where deviation from the light-cone is characterized by rapidity factors: $y_A$ and $y_B$. The limit where $y_A\to\infty$ and $y_B\to-\infty$ recovers the light-like directional vectors. The Collins soft function at large rapidity
then has the renormalized form \cite{Collins:2011zzd}:
\begin{align}
    \begin{split}
    S_Q\left(b_\perp; \Delta y; \mu\right) 
    =
    S_I\left(b_\perp, \mu\right) e^{\gamma_q \,\, \Delta y}
    +\mathcal O \left(e^{-2\Delta y }\right) \, ,
    \end{split}
    \label{eq:collins_soft}
\end{align}
where $S_I$ is the intrinsic soft factor ~\cite{Ji:2019sxk}, and $\Delta y = y_A - y_B$.

We may write the Euclidean analogue to $n_A$ and $n_B$ with purely imaginary Euclidean time components:
\begin{align}
\tilde n_A &= (0, 0, n_A^3, in_A^0) \, , \quad
\tilde n_B = (0, 0, -n_B^3, in_B^0) \, .
\label{eq:eucl_vec}
\end{align}
Our one-loop perturbative calculation indicates that
the soft function in Euclidean space with complex directional vectors is equivalent to its Minkowskian counterpart with real directional vectors. 
This equivalence requires that the
Euclidean directional vectors be mapped to the rapidity parameters in Minkowski space as (see \AppOne ):
\begin{align}
  r_a = \frac{n_A^3}{n_A^0} = \frac{1+e^{-2y_A}}{1-e^{-2y_A}}
  \, , \quad 
  r_b = \frac{n_B^3}{n_B^0} = \frac{1+e^{2y_B}}{1-e^{2y_B}} \, .
  \label{eq:spaceR}
\end{align}
Furthermore, the above equivalence can only be established 
with the conditions, 
$|r_a|,|r_b|>1$ and $n_A^0 n_B^0 (r_a r_b +1) >0$,
which corresponds to the space-like regime.
Such analyticity was discussed in detail in
\cite{Liu:2022nnk}. 

Under lattice regularization Wilson lines introduce power divergences in $\ell/a$, where $a$ is the lattice spacing (see \AppOne).
After subtracting them, the result approaches the soft function in the $\ell\to\infty$ limit.

\paragraph{Strategy for extracting the CS kernel.}

\begin{figure}
    \includegraphics[scale=0.19]{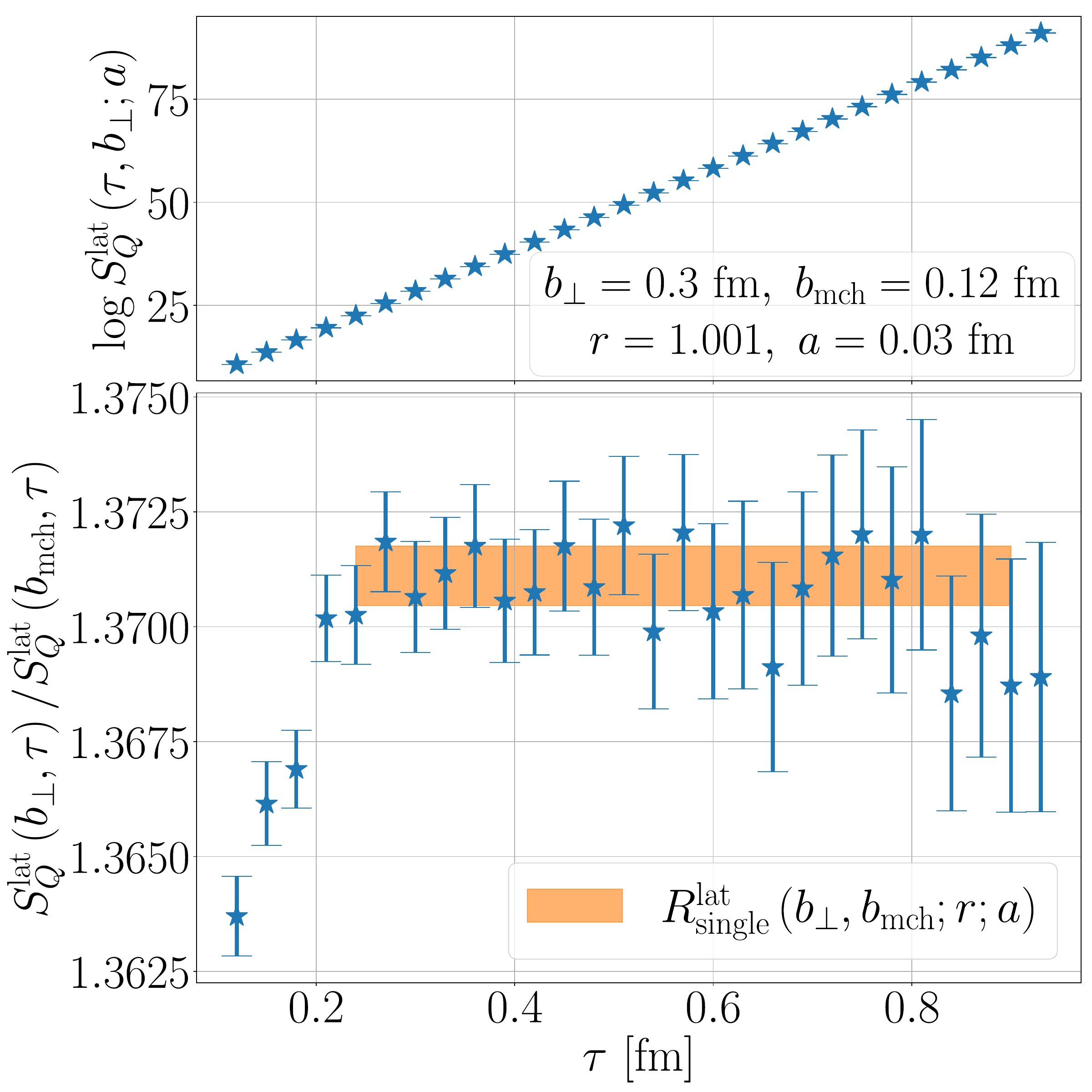}
    \caption{Time dependence of $S_{Q}^{\rm lat}$ and the ratio in Eq.~(\ref{eq:single}).
    }
    \label{fig:bfplat}
\end{figure}

\begin{figure*}
    \includegraphics[scale=0.21]{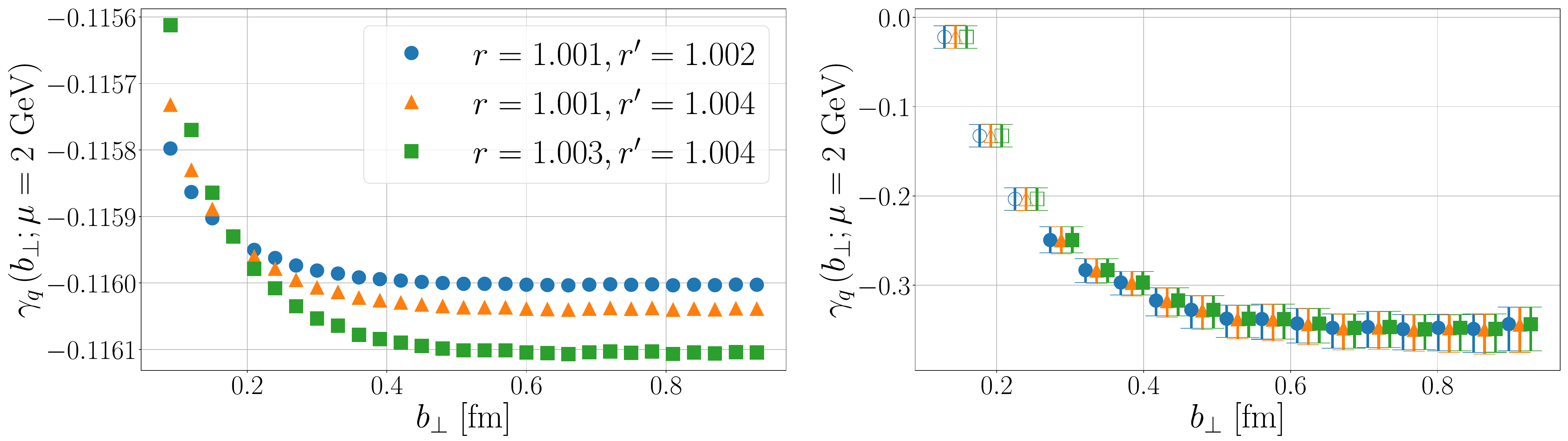}
    \caption{
    The CS kernel determined on the $a=0.03$ fm ensemble with three sets of bare rapidities.  Left: result obtained with Eq~(\ref{eq:Rdouble_factorize}) and matching  $\gamma_{q}  = \gamma_{q}^{\overline{\rm MS}}$ at $b_{\perp} = 0.18$ fm, using bare rapidities to evaluate $\Delta Y$.  Right: extracted CS kernel through Eq.~(\ref{eq:pfit}).  Open points are from dividing log($R_{\rm double}^{\rm lat}$) by $\Delta Y_{i}^{\rm ren}$, followed by the subtraction of the $c_{1}$ term, in the region $[b_{\rm cut}, b_{\rm th}]$.  Solid symbols represent $d_{l} + \gamma_{q}^{\rm mch}$. Points are shifted horizontally for visibility.
}
    \label{fig:rcomp}
\end{figure*}

We write the Wilson line in terms of a one-dimensional auxiliary 
fermion 
field that propagates along its path.
\cite{Arefeva:1980zd, Gervais:1979fv}.
The Green function, $H_n(x-y)$, associated with these auxiliary fields is obtained by solving the equation:
\begin{align}
  in\cdot D H_n(x-y) = i\delta^{(4)}(x-y) \, ,
  \label{eq:grM}
\end{align}
with covariant derivative, $D^\mu=\partial^\mu+ig_0 A^\mu$, and auxiliary field propagator $H_n$. 
After performing a Wick rotation, the Euclidean space Green function is:
\begin{align}
  i \tilde n \cdot D_E H_{\tilde n} (x_E-y_E) = \delta^{(4)}(x_E-y_E) \, .
  \label{eq:greenE}
\end{align}
The directional vector, $\tilde n$, now has a purely
imaginary time component, and can be written as $\tilde n =
(\vec n,in^0)$ in terms of the components of the Minkowski
space vector. 
To implement this on the lattice we approach the problem from the point of view of lattice NRQCD \cite{Thacker:1990bm} and its generalization to moving frames \cite{Mandula:1990fit,Horgan:2009}, which is equivalent to HQET in the infinite mass limit. 
The lattice auxiliary propagator can be computed by solving the lattice analogue to Eq. (\ref{eq:greenE}):
\begin{align}
\begin{split}
H_{\tilde n}( \tau)
&=
\Big(1-(2j)^{-1}\,\vec n \cdot \vec D_E (\tau)\Big)^j
U^\dag_4 \left(\tau-1\right)
\\
& \times \Big(1-(2j)^{-1}\,\vec n \cdot \vec D_E(\tau -1)\Big)^j
H_{\tilde n}\left( \tau-1\right)
,
\end{split}
\end{align}
where the small integer $j$ improves numerical stability.
 
To extract the Collins soft function defined in Eqs. (\ref{eq:soft_def}) and (\ref{eq:collins_soft}), we start by computing
\begin{align}
    \begin{split}
    & S_{Q}^{\rm lat}\left(b_\perp, \tau;  r_a, r_b; a\right) \\
    & \quad =
        \frac 1{N_c} \bra{0} \Tr
        H_{\tilde n_A}\left(0; \tau\right)
        W_{\hat b_\perp}\left(\tau\tilde n_A; 0,b_\perp\right) \\
            & \qquad\qquad \times
        H_{\tilde n_A}\left(b_\perp; -\tau\right)
        H_{\tilde n_B}\left(b_\perp; \tau\right) \\
            & \qquad\qquad \times
        W_{\hat b_\perp}\left(\tau\tilde n_B; b_\perp, 0\right)
        H_{\tilde n_B}\left(0; -\tau\right)
    \ket{0} \, ,
    \end{split}
    \label{eq:bfly_op}
\end{align}
drawn schematically in Fig. \ref{fig:bflyL}.

Equations (\ref{eq:greenE}) and
(\ref{eq:bfly_op}) suffer from UV cutoff effects also in the case of the free propagator. 
These cutoff effects have been investigated in the context of moving heavy quark effective theory 
\cite{Aglietti:1992in, Aglietti:1993hf}, 
and we expect that the most severe will appear as $e^{c\tau/a}$ in $S_{Q}^{\rm lat}$, where $c$ is a constant. 
To remove this exponential dependence on the cutoff we construct a single ratio,
\begin{align}
    R_{\rm single}^{\rm lat}\left(b_\perp, b'_\perp; r; a\right) = 
    \lim_{\tau \to \infty}
    \frac{S_Q^{\rm lat}
    \left(b_\perp, \tau; r, r; a\right)}{S_Q^{\rm lat}\left(b'_\perp, \tau; r, r;
  a\right)} \, .
    \label{eq:single}
\end{align}
This ratio is free of
 logarithmic divergences, but still suffers from power divergences in the form of $\left(b_\perp/a\right)^n$. 
In the large $\tau$ regime $S_Q^{\rm lat}(b_\perp,\tau)/S_Q^{\rm lat}(b'_\perp,\tau)$ is then expected to exhibit a plateau. In Fig. \ref{fig:bfplat} we demonstrate the $e^{c\tau/a}$ behavior in $S_Q^{\rm lat}$ and the appearance of this large $\tau$ plateau, which we then fit in order to determine $R_{\rm single}^{\rm lat}$. 
To remove the aforementioned power divergence in this ratio we construct a \textit{double ratio}:
\begin{align}
    R_{\rm double}^{\rm lat} \left(b_\perp, b'_\perp; r, r'\right) 
    &= \frac{R_{\rm single}^{\rm lat}\left(b_\perp, b'_\perp; r; a\right)}
      {R_{\rm single}^{\rm lat}\left(b_\perp, b'_\perp; r'; a\right)}
    \, ,
    \label{eq:ratioD}
\end{align}
which does not contain any UV divergence.  $R_{\rm double}^{\rm lat}$ still needs to be matched to its continuum counterpart through finite renormalization.  Using Eq. (\ref{eq:collins_soft}), one can show that (modulo lattice artifacts)
\begin{equation}
\label{eq:Rdouble_factorize}
R_{\rm double}^{\rm lat} \left(b_\perp, b'_\perp; r, r'\right)  = e^{\Delta \gamma_{q}(b_{\perp},b_{\perp}^{\prime}) \Delta Y (r,r^{\prime})}
\, ,
\end{equation}
with $\Delta\gamma_{q} (b_{\perp}, b_{\perp}^{\prime})$ being the difference between the CS kernel evaluated at $b_{\perp}$ and $b_{\perp}^{\prime}$, and  
\begin{align}
    \Delta Y \left(r, r'\right) 
    &= 
    2 \log\left(\frac{r-1}{r+1} \middle
        / \frac{r'-1}{r'+1}\right) \, .
\end{align}
Because $\Delta\gamma_{q}$ is renormalization group invariant, matching $R_{\rm double}^{\rm lat}$ to its continuum counterpart amounts to the renormalization of $r$ and $r^{\prime}$, which is required on the lattice because $O(4)$ space-time symmetry is broken~\cite{Aglietti:1993hf}.

The investigation of the double ratio only allows us to extract $\Delta \gamma_q (b_{\perp}, b_{\perp}^{\prime})$ from lattice data.  To determine $\gamma_q\left(b_\perp^{\prime}, \mu\right)$, we
match our calculation to the perturbative value of 
$\gamma_q\left( b_\perp, \mu \right)$ 
at $b_\perp \ll 1/\Lambda_{\rm QCD}$. For this purpose, we employ the $\rm \overline{MS}$ results at 
next-to-next-to-next-to-leading logarithmic order (N$^3$LL).
\cite{Li:2016ctv,Vladimirov:2016dll,Korchemsky:1987wg,Moch:2004pa,Henn:2019swt,vonManteuffel:2020vjv,Moult:2022xzt,Duhr:2022yyp}.  Furthermore, we use additional data points in the perturbative regime to obtain the effect of the rapidity renormalization on $\Delta Y$.

\paragraph{Numerical implementation and results.}

Our computations are performed on three quenched Wilson gauge ensembles  with $L_{\rm phys}\simeq 2~\rm{fm}$, where
$a \in \{0.048,\, 0.041,\, 0.03\} ~{\rm fm}$,
$L \in \{ 40,\, 48,\, 64\}$, $T = 2L$ and $N_{\rm config} \in \{250,\, 341,\, 200\}$ \cite{Detmold:2018zgk}. 
With $2048$ sources per configuration we achieve sub-percent statistical precision.

On all ensembles, we choose the same bare rapidities, 
$r,r'\in\left\{1.001, 1.002, 1.003, 1.004\right\}$, and 
compute $S_Q^{\rm lat}$ with
$\tau, b_\perp \in \left[0,(L/2-1)a\right]$.  
Using data for $S_Q^{\rm lat}$, we construct $R_{\rm double}^{\rm lat}$, and fit it with the ansatz,
\begin{align}
    &\log R_{\rm double}^{\rm lat} 
    \left ( b_{\perp}, b_{\rm mch}; r, r^{\prime}\right ) = 
    \Delta Y^{\rm ren}_{i} \nonumber \\
    &\quad \times \left(
        \left\{ \begin{array}{lr}
            \gamma_{q}^{\rm \overline{MS}} \left(b_\perp, \mu\right) - \gamma_q^{\rm mch}
             &,~
                b_{\rm cut} \leq b_\perp \leq b_{\rm th} \\
            \left\{d_\ell \right\} &,~~~
                b_\ell = b_\perp > b_{\rm th}
        \end{array} \right\}
        \right. \nonumber \\ &\qquad \left.
     + c_1 \left( \frac{a^2}{b_\perp^2} - \frac{a^2}{b_{\rm mch}^2} \right)
    \right) \, ,
    \label{eq:pfit}
\end{align}
where $\Delta Y^{\rm ren}_{i}$ is the renormalized $\Delta Y$ on ensemble $i$, $\gamma_{q}^{\rm \overline{MS}}$ is the $\overline{\rm MS}$-scheme perturbative result evaluated at the renormalization scale $1/b_\perp$ and evolved to a common $\mu = 2~{\rm GeV}$ 
at ${\rm N}^{3}$LL~
\cite{Li:2016ctv,Vladimirov:2016dll,Korchemsky:1987wg,Moch:2004pa,Henn:2019swt,vonManteuffel:2020vjv,Moult:2022xzt,Duhr:2022yyp}, $b_{\rm mch}$ is a chosen parameter, 
$b_{\rm th}$ separates the perturbative and non-perturbative regions in $b_\perp$, and data at $b_\perp < b_{\rm cut}$ are excluded in the fit because of large lattice artifacts. 

In Eq.~(\ref{eq:pfit}), $\Delta Y_{i}^{\rm ren}$, $\gamma_{q}^{\rm mch} \equiv \gamma_{q}(b_{\rm mch}, \mu)$, $d_{l}$ and $c_{1}$ are fit parameters. 
This equation represents a global fitting procedure that uses data in the perturbative region, $\left[b_{\rm cut}, b_{\rm th}\right]$, to extract the value of $\gamma_{q}^{\rm mch}$ and gives results of $\Delta \gamma_{q}(b_{\perp}, b_{\rm mch})$ (the $d_{l}$ parameters) in the non-perturbative regime.  
We also parameterize order $a^2$ cutoff effects with the $c_1$ term. 
There can be additional lattice artifacts that we will estimate as a systematic error in our numerical calculation explained below.

To minimize lattice artifacts while staying in the perturbative region, we choose $b_{\rm cut} = 0.144~{\rm fm}$, and $b_{\rm th} = 0.24~{\rm fm}$, corresponding to three and five lattice spacings of our coarsest ensemble.
We work with $b_{\rm mch}$ outside the perturbative region, and find that $b_{\rm mch} = 0.384~{\rm fm}$ produces the best fit. 

A correlated fit, using Eq. (\ref{eq:pfit}), results in a large 
$\chi^2/N_{\rm dof}$.
This we attribute to
the high 
statistical precision in our method, which leads to 
prominent enhancement of $\chi^2$ from any systematic error. 
We find that it is mainly caused 
by the discretization effects that are not subtracted by the 
$c_1$ term. 
These effects can be 
regarded as the systematic error arising from the continuum 
extrapolation.
We estimate their size by performing the analysis omitting one ensemble at a time.
More details can be found in \AppTwo. 

Fitting at several pairs of $(r,r')$ allows us to check the consistency of our result, since the CS kernel is independent of the rapidity. 
Figure \ref{fig:rcomp} shows that all renormalized rapidities give 
consistent results for the CS kernel, 
while the bare ones fail to do so. This demonstrates the 
importance of the rapidity renormalization. 
It also indicates that our result is well described by Eq. (\ref{eq:collins_soft}), leading to definitive evidence for the validity of our approach.

\begin{figure}
    \includegraphics[scale=0.21]{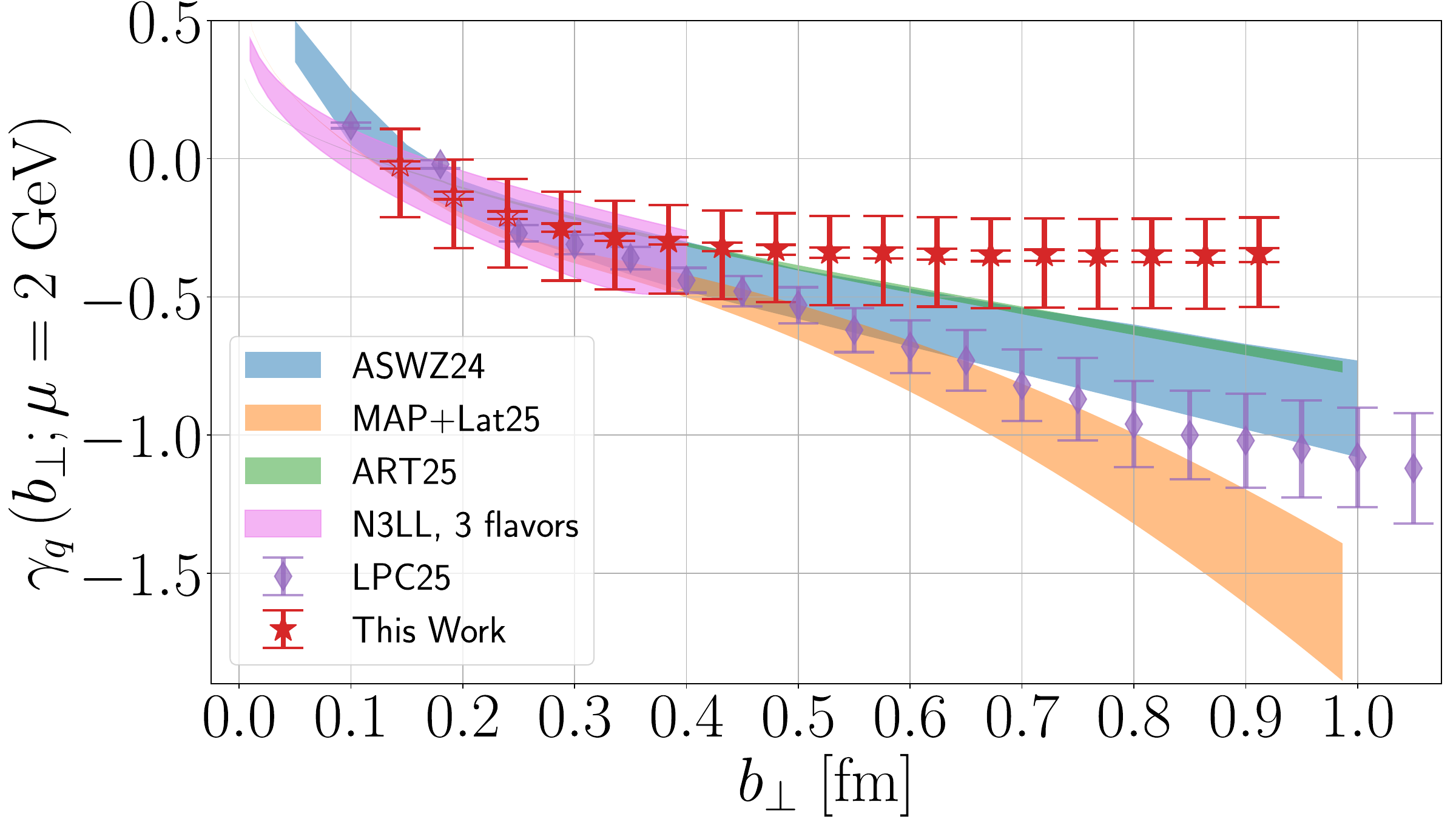}
    \caption{
    Our result for the CS kernel compared with other recent
    lattice and phenomenological extractions. 
    Filled stars represent $d_\ell+\gamma^{\rm mch}_q$.
    Open stars are from dividing log($R_{\rm double}^{\rm lat}$) at $a=0.03$ by $\Delta Y_{0.03}^{\rm ren}$, followed by subtracting the $c_{1}$ term, in the perturbative region $[b_{\rm cut}, b_{\rm th}]$.
    The data from the other two ensembles are compatible in this region.
    The inner error bars include statistical, truncation and 
    discretization error. The outer error bars also include 
    the scale variation error with factor $\sqrt{2}$.
    Recent lattice extractions are: 
    ASWZ24 \cite{Avkhadiev:2024mgd}, 
    LPC25 \cite{Tan:2025ofx}
    A recent phenomenological extraction is:
    ART25 \cite{Moos:2025sal}. 
    MAP+Lat25 \cite{Avkhadiev:2025wps} is a combined analysis using lattice and experimental data.
    }
    \label{fig:kcs_final}
\end{figure}

Our error budget includes contributions from statistical uncertainties, 
perturbative series truncation, and lattice artifacts. 
We estimate the truncation error by comparing 
results 
using 
N$^3$LL or next-to-next-to-leading logarithmically (N$^2$LL) resummed CS kernels in the fits.
For the extracted CS kernel,
the relative statistical
and truncation errors are respectively $\sim 0.46\%$ and $\sim 0.74\%$ on average, while 
the discretization error averages to be between
$\sim 4.8\%$ and $\sim 6.5\%$.
In addition, the uncertainty of matching in the perturbative region is estimated by varying $\mu$ by a factor of $\sqrt{2}$, which constitutes the dominant source of systematic error
(see \AppTwo~ for details). 
Our result for the CS kernel is given in Fig. \ref{fig:kcs_final}, where the larger error bar accounts for all aforementioned errors taken in quadrature, while the smaller one omits the matching uncertainty. 

This calculation is performed in the quenched approximation, allowing us to access ensembles at fine lattice spacings.
We find the quenching error is $\sim 15\%$ 
by comparing $R_{\rm double}^{\rm lat}$ obtained from quenched and dynamical simulations at a coarser lattice spacing, $a\simeq 0.09~{\rm fm}$, with $L=32$, $T=2L$ using the Iwasaki gauge action, with dynamical Wilson-Clover fermions at light quark mass, $m_{\rm ud} = 2.527(47)~{\rm MeV}$, 
and strange quark mass, $m_{\rm s} = 72.72(78)~{\rm MeV}$ \cite{PACS-CS:2008bkb}.
This implies that quenching is not currently the dominant systematic effect.

Figure \ref{fig:kcs_final} exhibits that the CS kernel from our calculation plateaus in the region $b_{\perp} \gtrsim 0.5$ fm.  This is consistent with the prediction from~\cite{Collins:2014jpa}, that the CS kernel should be constant at large $b_{\perp}$.
While our result 
is consistent with other lattice computations at intermediate $b_\perp$, it disagrees for $b_{\perp} \gtrsim 0.5$ fm. 
It is interesting to see if this plateau behavior persists in dynamical computations.

\paragraph{Conclusion.}

We develop a method for computing the 
CS kernel in lattice QCD from a vacuum soft function. 
This method does not require the choice of any external hadronic state. 
We also present the first numerical results in the quenched approximation with a continuum extrapolation.  Our approach benefits from high statistical precision and low computational cost, with
the dominant systematic effect being the uncertainty in matching to perturbative results.
Errors from the continuum extrapolation are also non-negligible.
Nevertheless, there is a clear route toward future improvement of the systematic uncertainties, including simulating at finer lattices and developing an alternative scheme for the rapidity renormalization.  
This will result in 
a precise determination of the CS kernel, which will be crucial for phenomenological extractions of TMDPDFs, as relevant to physics of the Electron-Ion Collider. 
Finally, the Wilson lines we consider on the lattice may have relevance to small-$x$ and heavy ion physics given their interpretation as Eikonal propagators \cite{McLerran:1993ni,Liu:2006ug,Rajagopal:2025rxr,Mehtar-Tani:2025xxd}.

{\sl Acknowledgments ---} The authors are grateful to William Detmold and Mike Endres for providing us with the 
quenched ensembles used in this project.  We thank William Detmold and Issaku Kanamori for their participation in the early stage of this work.
We would also like to acknowledge Yi-Zhuang
Liu for interesting and helpful discussions.
Numerical calculations in this work were carried out on an Intel-KNL cluster at National Yang Ming Chiao Tung University.  We thank ASRocK Inc. for their support of the construction of this cluster.  The work of AF is supported by grant numbers NSTC113-2112-M-A49-018-MY3 and NSTC111-2112-M-A49-018-MY2.
CJDL and WM acknowledge the support from Taiwanese NSTC through grant numbers 112-2112-M-A49-021-M03, 114-2123-M-A49-001-ASP and 114-2811-M-A49-516. The work of YZ is supported by the U.S. Department of Energy, Office of Science, Office of Nuclear Physics through Contract No.~DE-AC02-06CH11357, and the Early Career Award through Contract No.~DE-SCL0000017.


\bibliographystyle{apsrev4-1}
\bibliography{refs} 
\clearpage
\appendix

\section{End Matter}
\paragraph{Appendix A: Perturbative results.} 
\label{app:perturbative_results}

\begin{figure}
    \def\ww{0.73}
    \subfloat[]{\includegraphics[scale=\ww]{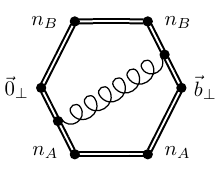}} 
    \subfloat[]{\includegraphics[scale=\ww]{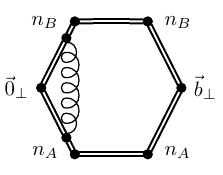}} 
    \subfloat[]{\includegraphics[scale=\ww]{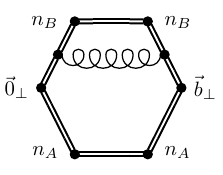}} \\
    \subfloat[]{\includegraphics[scale=\ww]{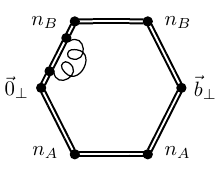}} 
    \subfloat[]{\includegraphics[scale=\ww]{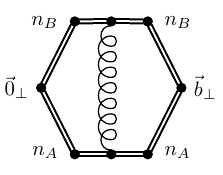} \label{fig:feyn:e}}
    \subfloat[]{\includegraphics[scale=\ww]{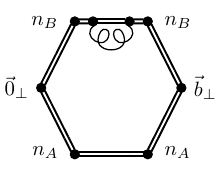} 
    \label{fig:feyn:f}}
    \caption{Feynman diagrams relevant at one-loop in perturbation theory. Double lines represent Wilson lines, and curly lines represent gluons.}
    \label{fig:feyn}
\end{figure}

To gain insight into the lattice computation, we carry out an $\mathcal{O}(\alpha_{s})$ perturbative computation 
in Euclidean space with finite length Wilson lines, using the Polyakov regulator \cite{Polyakov:1980ca}, $a_P$, 
to regulate the UV divergence. 
Performing a large-$\ell$ expansion, the results of the perturbative computations are:


\begin{align}
\begin{split}
    &S_{(a)}\left(b_\perp, \ell; n_A, n_B; a_{P}\right) + {n_A \leftrightarrow n_B}
    \\ 
    &~= \frac{\alpha_{s} C_F}{\pi}  
    \frac{R_{ab}}{2}
    \mathbf{L}_b  \sum_{i=a,b}\mathbf{L}^{(1)}_{i} 
    + C_A
    + \mathcal O \left(\frac{b_\perp^2}{\ell^2}\right) \, ,
\end{split}
\\
\begin{split}
    &S_{(b)}\left(b_\perp, \ell; n_A, n_B; a_{P}\right) + {n_A \leftrightarrow n_B}
    \\ 
    &~= -\frac{\alpha_{s} C_F}{\pi}  
\frac{R_{ab}}{2}
    \mathbf{L}_a  \sum_{i=a,b}\mathbf{L}^{(1)}_{i} 
     - C_A
    + \mathcal O \left(\frac{a_P^2}{\ell^2}\right) \, ,
\end{split} \\
\begin{split}
    &S_{(c)}\left(b_\perp, L; n_A, n_B; a_{P}\right) + {n_A \leftrightarrow n_B}
    \\ 
    &~= \frac{\alpha_{s} C_F}{\pi}  
    \left(
        -\mathbf{L}_b
        +\frac{\pi \ell }{b_\perp}  R^{(2)}
    \right)+ C_B
    + \mathcal O \left(\frac{b_\perp^2}{\ell^2}\right) \, ,
    \label{eq:pertc}
\end{split} 
\\
\begin{split}
    &2S_{(d)}\left(b_\perp, \ell; n_A, n_B; a_{P}\right) + {n_A \leftrightarrow n_B}
    \\ 
    &~= \frac{\alpha_{s} C_F}{\pi}  
    \left(
        \mathbf{L}_a
        -\frac{\pi \ell }{a_P} R^{(2)}
    \right) - C_B
    + \mathcal O \left(\frac{a_P^2}{\ell^2}\right) \, ,
\end{split} \\
\begin{split}
    &S_{(f)}\left(b_\perp, \ell; n_A, n_B; a_{P}\right)
    \\ 
    &~= \frac{\alpha_{s} C_F}{\pi} 
    \left(
        \log \left(\frac{b_\perp^2}{a_P^2}+1\right)
        -\frac{ 2b_\perp \arctan\left(\frac{b_\perp}{a_P}\right)}{a_P}
    \right) \, ,
\end{split}
\end{align}
where $\mathbf{L}^{(1)}_{i} = \log \left ( (r_{i}-1)/(r_{i}+1) \right )$, $\mathbf{L}_b = \log\left(\ell^2/b_\perp^2\right)$ and
$\mathbf{L}_a = \log\left(\ell^2/a_P^2\right)$, $R_{ab} = (r_{a}r_{b}+1)/(r_{a}+r_{b})$, $R^{(2)}=\sum_{i=a,b}\sqrt{(r^{2}_{i}-1)/(r^{2}_{i}+1)}$. 
Also, $C_{A/B}$ are constants in $\ell$, $b_\perp$, and $a_P$ that result from the large-$\ell$ expansion, and cancel in the sum of diagrams. Finally, $C_F$ is a QCD color factor.
The first four diagrams in Fig.~\ref{fig:feyn}, with their 
counterparts of $n_A\leftrightarrow n_B$, contain their infinite-$\ell$ analogues.  Linear divergences in $\ell$ appear in $S_{(c,d)}$.  They are not canceled in the double ratio in Eq.~(\ref{eq:ratioD}), but are suppressed when $r_{a,b}$ are close to 1.
Figure (\ref{fig:feyn:e}) is of $\mathcal{O}(b^{2}_{\perp}/\ell^{2})$  and vanishes when $\ell \rightarrow \infty$. 
Figure (\ref{fig:feyn:f}) is $\ell$-independent, since it is just the self-energy of the transverse Wilson-line, but it does contribute a linear divergence at $a_{P}\rightarrow 0$, 
which cancels in the double ratio.
Using Eq. (\ref{eq:spaceR}) to rewrite $r_{a/b}$ in terms of $y_{A/B}$, we can see that up to linear divergences and power corrections, we are able to reproduce the one-loop result for the soft factor in Minkowski space. See Eqs. (B11-B13) in \cite{Ebert:2019okf} for comparison.

\paragraph{ Appendix B: Numerical results in detail.}
\label{app:numerical_details} 

\begin{table*}
    \begin{tabular}{ c*{19}{c} } 
    \hline 
    \hline 
        & 
        $\vbox{\vspace{3mm}}\Delta Y_{0.048}$ & $\Delta Y_{0.041}$ & $\Delta Y_{0.03}$ & $c_1$ & $\gamma_q^{\rm mch}$ \\
    \hline 
        result & 
        $\vbox{\vspace{3mm}}-2.15\times 10^{-4}$ & $-2.10\times 10^{-4}$ & $-1.69\times 10^{-4}$ & $3.25$ & $-3.09\times 10^{-1}$ \\
        $\vbox{\vspace{3mm}}\sigma_{\rm stat}$ &
        $7.47\times 10^{-6}$ & $6.37\times 10^{-6}$ & $3.88\times 10^{-6}$ & $1.92\times 10^{-1}$ & $1.93\times 10^{-3}$ \\
        $\vbox{\vspace{3mm}}\sigma_{\rm trunc}$ &
        $2.43\times 10^{-6}$ & $2.24\times 10^{-6}$ & $1.62\times 10^{-6}$ & $4.94\times 10^{-2}$ & $2.56\times 10^{-3}$ \\
        $\vbox{\vspace{3mm}}\sigma^+_{\rm disc}$ &
        $1.38\times 10^{-4}$ & $3.43\times 10^{-5}$ & $5.89\times 10^{-5}$ & $1.31$ & $1.21\times 10^{-2}$ \\
        $\vbox{\vspace{3mm}}\sigma^-_{\rm disc}$ &
        $5.60\times 10^{-5}$ & $5.16\times 10^{-5}$ & $1.98\times 10^{-5}$ & $1.78$ & $1.37\times 10^{-2}$ \\
    $\vbox{\vspace{3mm}}\sigma^+_{1.2}$ &
        $ 4.21\times 10^{-5}$ & $ 3.96\times 10^{-5}$ & $ 2.97\times 10^{-5}$ & $6.34\times 10^{-1}$ & $7.50\times 10^{-2}$ \\
    $\vbox{\vspace{3mm}}\sigma^-_{1.2}$ &
        $ 4.15\times 10^{-5}$ & $ 3.93\times 10^{-5}$ & $ 2.98\times 10^{-5}$ & $9.61\times 10^{-1}$ & $9.17\times 10^{-2}$ \\
    $\vbox{\vspace{3mm}}\sigma^+_{\sqrt{2}}$ &
        $ 8.02\times 10^{-5}$ & $ 7.52\times 10^{-5}$ & $ 5.63\times 10^{-5}$ & $1.04$ & $1.33\times 10^{-1}$ \\
    $\vbox{\vspace{3mm}}\sigma^-_{\sqrt{2}}$ &
        $ 7.80\times 10^{-5}$ & $ 7.41\times 10^{-5}$ & $ 5.65\times 10^{-5}$ & $2.33$ & $1.96\times 10^{-1}$ \\
    \hline
    \hline 
    \end{tabular} 
    \caption{Results of fitting procedure with estimated uncertainty.
    The $\pm$ superscripts indicate upper and lower errors.
    Parameters $d_\ell$ are not listed, but their values can be inferred from 
    Fig. \ref{fig:rcomp}. }
    \label{tab:fit_results} 
\end{table*}

\begin{figure}
    \includegraphics[scale=0.19]{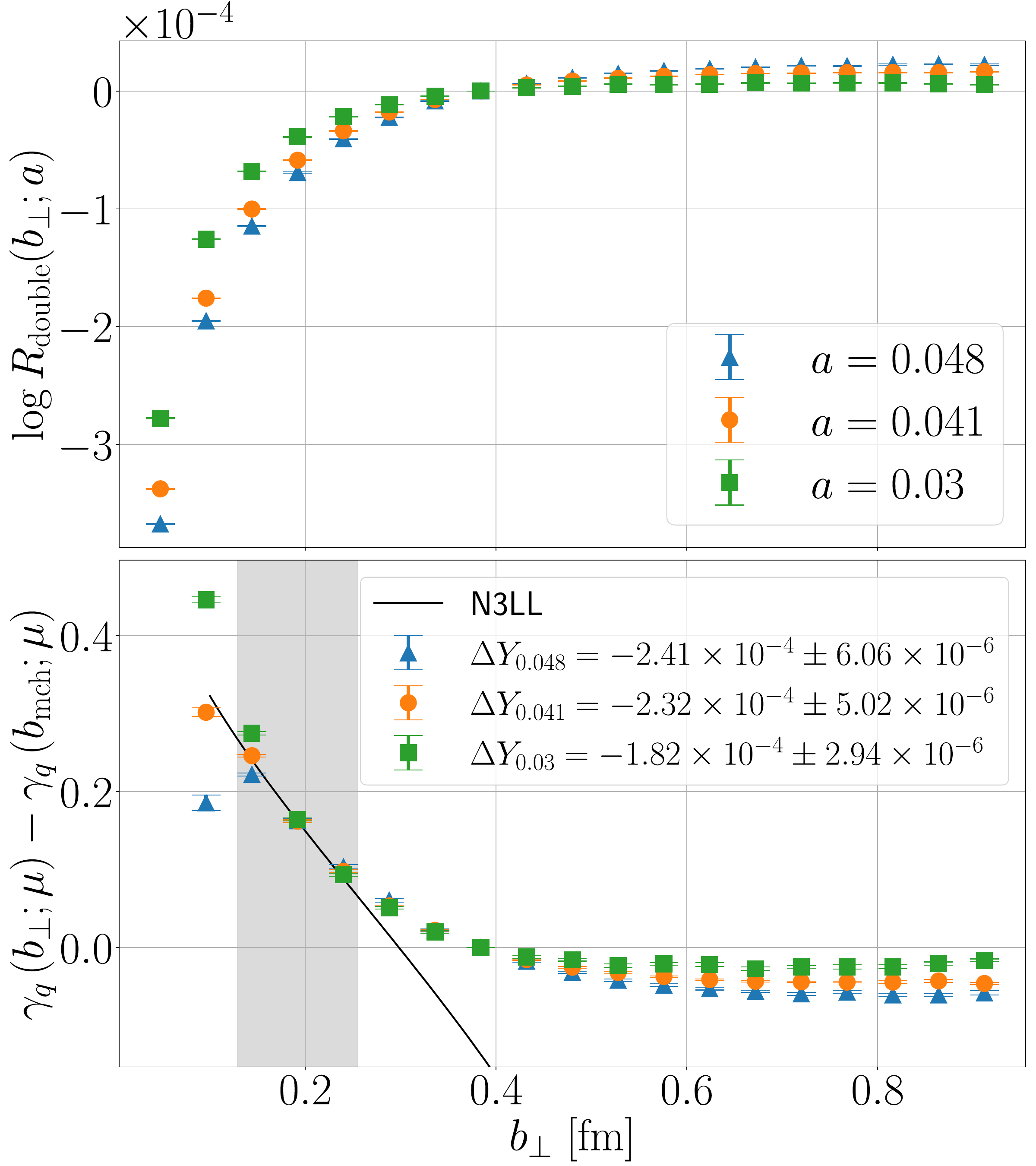}
    \caption{Top plot: Double ratio after interpolating data to common separation, $0.048~{\rm fm}$. Bottom plot: CS kernel difference obtained from individual data sets. Gray region highlights perturbative range used in fit: $b_{\rm min} \leq b_\perp \leq b_{\rm th}$}
    \label{fig:double_subtracted}
\end{figure}

\begin{figure}
    \includegraphics[scale=0.21]{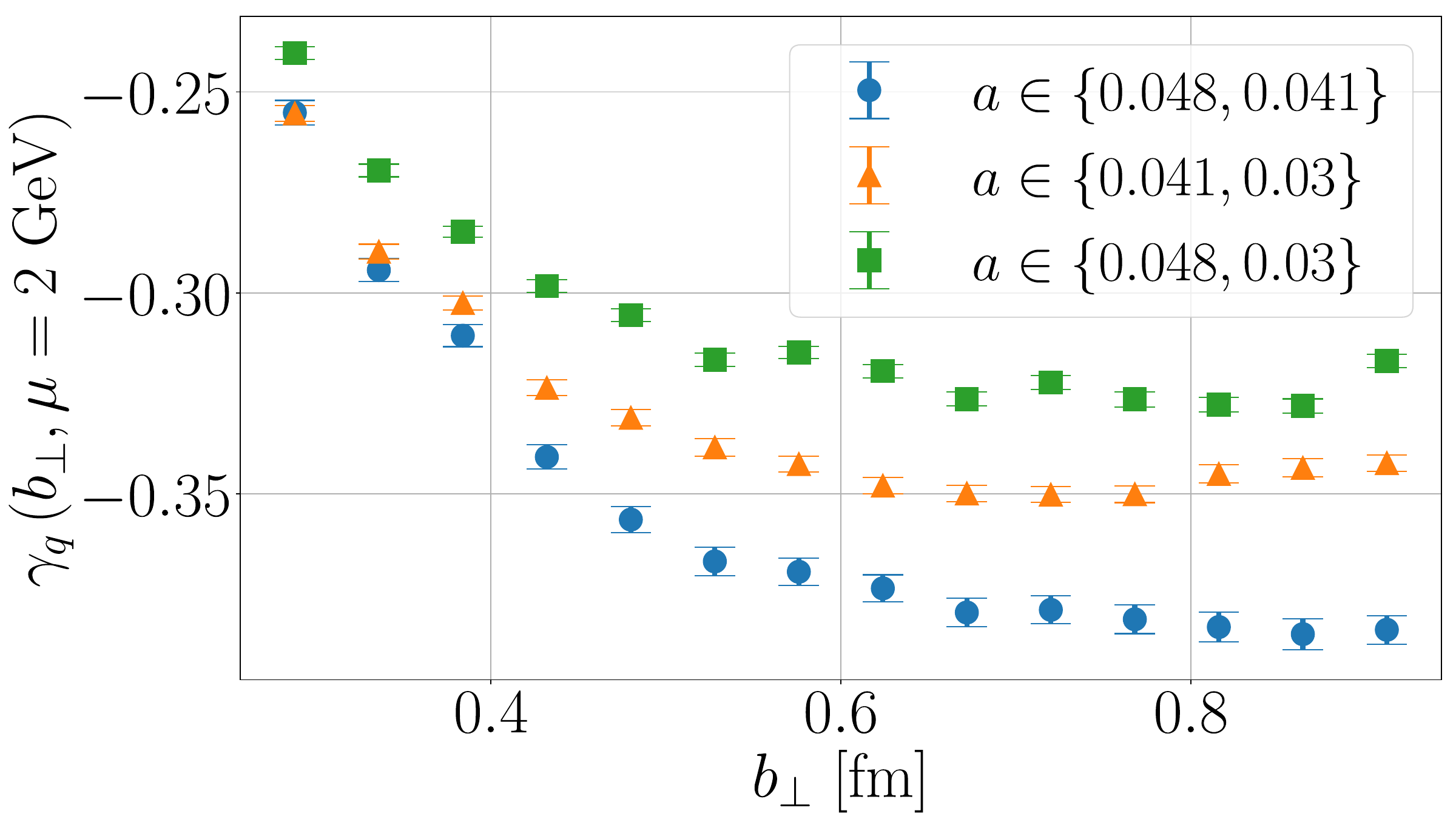}
    \caption{Extracting the CS kernel using two ensembles.}
    \label{fig:disc_plot}
\end{figure}

%
%
%

All sources of error from our analysis are listed in Tab. \ref{tab:fit_results}. The
statistical error $\sigma_{\rm stat}$ is obtained through a bootstrap analysis
of the data.
The truncation error 
$\sigma_{\rm trunc}$ is obtained by redoing the fit with the N$^2$LL result for
the $\overline{\rm MS}$ CS kernel and taking the difference with the N$^3$LL
fit.
The scale variation error is given by 
$\sigma^\pm_{s}$, with $s \in \{1.2, \sqrt{2}\}$.
The upper bound, with the $+$ superscript is obtained by varying the scale by
$s \mu$, while the lower bound, is obtained by varying the scale by
$\mu/s$.  
To estimate this error, we take the $\sqrt{2}$ variation as the outer error bars in Fig. \ref{fig:kcs_final}.

The top plot in Fig. \ref{fig:double_subtracted} presents 
the result of taking the double ratio after
interpolating the results to a common interval of $b_\perp=0.048~{\rm fm}$.
Lattice artifacts become more apparent further away from $b_{\rm mch}$. While our model attempts to account for the lattice artifacts in
the $c_1$ term, we see in the bottom plot in Fig. \ref{fig:double_subtracted}
that there is still lattice spacing dependence in the large $b_\perp$ region and the
small $b_\perp \lesssim b_{\rm min}$ region.

This discretization error is estimated by redoing the fit three separate
times, omitting one ensemble from the fit each time.  
The results, as shown in Fig. \ref{fig:disc_plot}, straddle that from the central analysis procedure of including all three ensembles.  The discretization errors are then obtained by comparing the central-procedure result with the upper and lower values in this figure.

\paragraph{Appendix C: A possible approach to obtain the soft function.}

\begin{figure}
    \vspace{4mm}
    \includegraphics[scale=0.26]{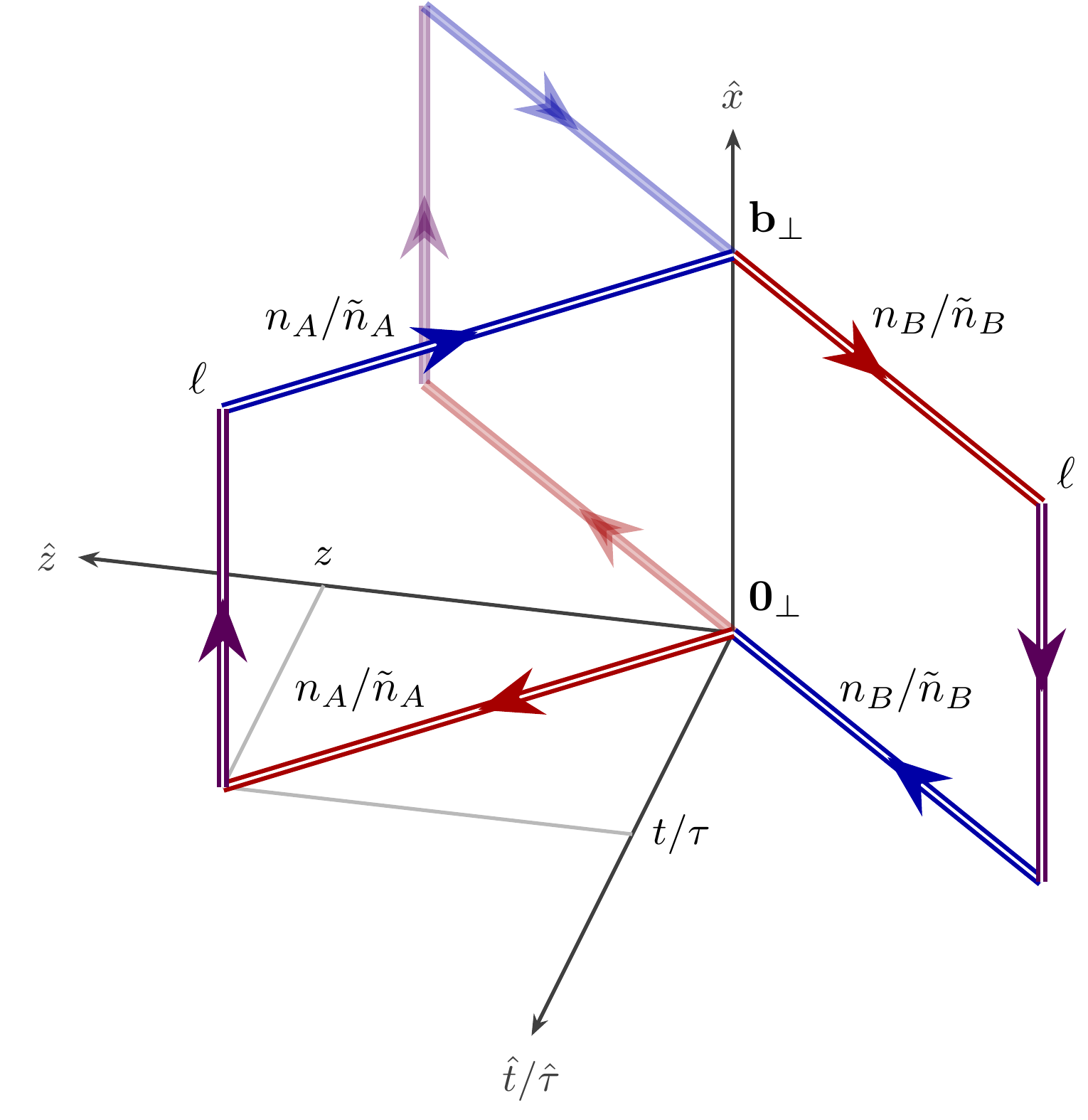}
    \caption{Same schematic as Fig. \ref{fig:bflyL}, but with rectangle loop, $S_{\rm rect}^{\rm lat}$, for $n_B/\tilde n_B$ represented with transparent staple.}
    \label{fig:bfly_rect}
\end{figure}

The cutoff effects, in addition to power divergences may be subtracted by taking ratio similar to that in ~\cite{Ji:2019sxk}:
\begin{align}
    &S^{\rm lat}_{\rm ratio}\left(b_\perp, \tau; r_a, r_b\right) \nonumber \\
    & \qquad = \frac{S_Q^{\rm lat} \left(b_\perp, \tau; r_a, r_b; a\right)}{\sqrt{S_{\rm rect}^{\rm lat}\left(b_\perp, \tau; r_a; a\right) S_{\rm rect}^{\rm lat}\left(b_\perp, \tau; r_b; a\right)}} \, ,
    \label{eq:ratio1}
\end{align}
which also handles the renormalization of the soft function.
The factor of $S_Q^{\rm lat}$ in the numerator is the same `butterfly' shaped object as before, while the denominator factors $S_{\rm rect}^{\rm lat}$ are rectangle shaped loops which are
represented schematically in Fig. \ref{fig:bfly_rect}.

In practice, measurements of  $S_{\rm rect}^{\rm lat}$ suffer from a signal to noise problem on the lattice.
This may be linked to the lattice artifacts that show up in
$S_{\rm rect}^{\rm lat}\sim e^{h_1 \tau /a + i h_2 z/a}$,
where $h_{1,2}$ are constants, and $z$ is the spatial distance traveled by the auxiliary field propagator in the rectangle loop.
In our numerical calculation, we do observe a rapidly-changing complex phase of $S_{\rm rect}^{\rm lat}$ across lattice configurations and source locations.

The oscillatory term in $S_{\rm rect}^{\rm lat}$ is related to the UV cutoff effects on the auxiliary propagator. In order to suppress these effects, we performed a computation of $S_{\rm rect}^{\rm lat}$ on gauge configurations after APE smearing. While this did produce a noticeable effect at early values of $\tau$, the oscillatory behavior dominated in the moderate to large-$\tau$ region.

Solving the noise problem in the rectangle loops allows for a direct computation of the Collins soft function in addition to the CS kernel, and is the subject of future work. See \cite{Morris:2025y} for more details on this method.

\end{document}